# Scaled Ultra-Wide Bandgap AlGaN Polarization-Graded FET with Ultra-thin Buffer Layer

Yinxuan Zhu[1, a)], Ashley Wissel-Garcia[4], Kidus Guye[6], Chandan Joishi[1], Can Cao[1], Seungheon Shin[1], Kyle Liddy[1,5], Emils G.B. Jurcik[6], Agnes Maneesha Dominic Merwin Xavier[1], Andrew A. Allerman[3], Brianna A. Klein[3], Andrew Amrstrong[3], James S. Speck[4], Samuel Graham[6], Siddharth Rajan[1, 2, a)]

[1] *Department of Electrical and Computer Engineering, The Ohio State University, Columbus, Ohio 43210, USA*

[2] *Department of Materials Science and Engineering, The Ohio State University, Columbus, Ohio 43210, USA*

[3] *Sandia National Laboratories, Albuquerque, New Mexico 87123, USA.*

[4] *Materials Department, University of California, Santa Barbara, California 93106, USA.*

[5] *Air Force Research Laboratory, Sensors Directorate, Wright-Patterson Air Force Base, Ohio 45433, USA*

[6] *Mechanical Engineering, University of Maryland, College Park, Maryland, USA*

**Abstract:**

We report on the design and demonstration of ultra-wide bandgap AlGaN polarization-graded field effect transistors with ultra-thin channels to enable excellent current density and high-frequency performance while significantly reducing thermal resistance. We use polarization-graded AlGaN layers and ultra-thin pseudomorphic AlGaN buffer layers to enable low thermal resistance and excellent structural quality. The polarization-graded field effect transistors (PolFETs) demonstrated here show $I_{max}$ over 800mA/mm and current/power gain cutoff frequency ($f_T/f_{max}$) of 26/28 GHz. Small signal modeling and analysis were used to determine parasitic/transit delays, and gate-resistance thermometry was implemented to thermally characterize AlGaN PolFET and benchmark against state-of-the-art AlGaN HEMTs. The ultra-thin AlGaN PolFET showed thermal resistance of 12 K·mm/W, representing a significant reduction from typical AlGaN transistors. These results show state-of-art combination of high current density, excellent $f_T$-$L_G$ product for ultra-wide bandgap AlGaN transistors, and superior thermal performance, and highlight the promise of AlGaN transistors for future RF and mm-wave applications.

[a)] Authors to whom correspondence should be addressed

Electronic mail: *zhu.2931@osu.edu*, *rajan.21@osu.edu*

# I. Introduction

Ultra-wide bandgap (UWBG) AlGaN lateral transistors are promising for high frequency power amplifier applications as their higher breakdown field leads to an improved Johnson Figure of Merit (JFOM) over GaN lateral transistors [1, 2]. To achieve high-performance UWBG AlGaN lateral transistors, several challenges must be addressed, including field management, current injection, and thermal management. High breakdown fields have already been demonstrated in UWBG AlGaN diodes (>8MV/cm) [3] and HFETs (>5MV/cm) [4]. High DC (current density > 1A/mm) [5] and RF performance ($f_T/f_{max}$ = 40 GHz/58 GHz) [6] have been demonstrated on UWBG AlGaN HEMTs with reverse graded contact layer in etched well or in-situ reverse graded contacts. To improve current injections, in-situ metal-organic chemical vapor deposition (MOCVD)-grown reverse graded contacts were used to achieve extremely low contact resistance on UWBG AlGaN in [7-8]. The higher anticipated operating voltage and power density in AlGaN transistors require good thermal management. The use of thin AlGaN layers (which have low thermal conductivity) and AlN heat spreading layers (in thin film or bulk wafer form) can enable low thermal resistance in high Al-composition AlGaN transistors. However, the growth of AlGaN channel layers on AlN leads to a negative polarization sheet charge due to the discontinuity in polarization can cause significant depletion and trapping effects. Prior work [2-6, 8-13] has typically used thick AlGaN buffer layers to transition from an AlN template to the AlGaN channel, leading to high thermal resistance since random-alloy AlGaN layers have very poor thermal conductivity [14]. Furthermore, the lattice mismatch constraints with the device design is an often-overlooked issue. Thick layers often show partial or complete stress relaxation, leading to poor morphology and structural quality. Therefore, from both the epitaxial growth quality and thermal management point of view, thin pseudomorphic buffers are important for future AlGaN transistors. In this work, we demonstrate integration of several critical concepts necessary to overcome the fundamental challenges of UWBG AlGaN devices – ultra-thin psuedomorphic AlGaN buffer growth, integration of thin buffers with lower thermal resistance, and polarization-graded channels to realize high current. This enables us to achieve polarization-graded field effect transistors (PolFETs) [13, 15-19] with current density ($I_{max}$) over 800mA/mm and current gain cutoff frequency ($f_T$)/maximum oscillation frequency($f_{max}$) of 26/28 GHz, which represent state-of-art combination of current density and $f_T$-$L_G$ product for UWBG AlGaN transistors, while enabling a reduced thermal resistance of 12 K·mm/W compared to thick AlGaN HEMTs.

## II. Experimental

The epitaxial design used here is shown in Fig. 1 (a), and consists of: a 300 nm AlN buffer; $2.5 \times 10^{20}$ cm$^{-3}$ Si delta doping; a 4 nm UID Al$_{0.5}$Ga$_{0.5}$N spacer to mitigate impurity scattering; a 40 nm graded channel layer (graded from 50% to 75% Al-content AlGaN) with Si concentration of $2 \times 10^{18}$ cm$^{-3}$. The total thickness of the pseudomorphic AlGaN layer is 45 nm. Si delta-doping can compensate for the negative polarization charge at the interface. Fig. 2 shows the band diagrams and charge profiles of the PolFET structure with different levels of Si delta doping. Low or no Si delta doping (Fig. 2(b)) at AlN/Al$_{0.5}$Ga$_{0.5}$N interface leads to uncompensated negative polarization charge at the interface, causing hole-trap induced current collapse observed in previous N-polar GaN HEMTs [20]. High Si delta doping at the interface (Fig. 2(d)) induces a parasitic electron channel. With the correct Si delta doping (Fig. 2(c)) during growth, we compensate for negative polarization charge and avoid the parasitic channel at the interface.

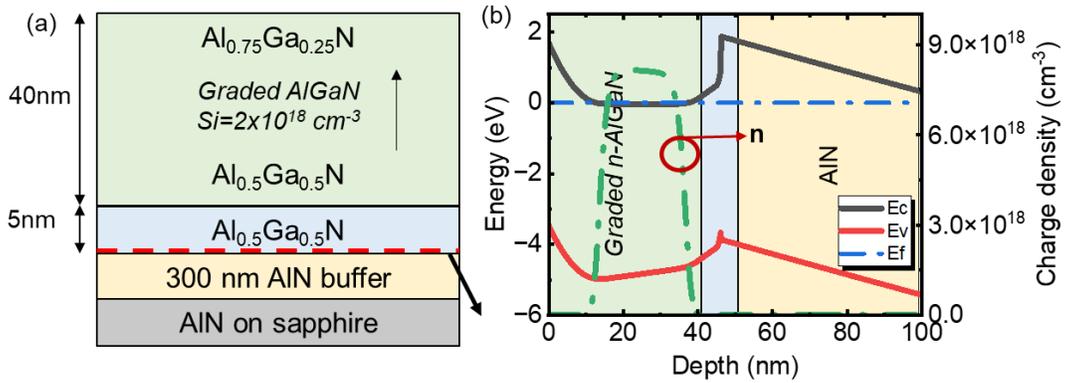

Fig. 1. (a) Epitaxial structure of Ammonia MBE-grown polarization-graded AlGaN FETs; (b) Band diagram of the epitaxial structure in (a) and charge profile generated by 1-D Schrodinger-Poisson solver [21].

The PolFET channel was grown on an AlN-on-sapphire template substrate consisting of approximately 3 μm AlN on a 0.2° miscut single-side-polished sapphire wafer. Epitaxial growth was done by ammonia molecular beam epitaxy (NH$_3$-MBE) with a Veeco Gen930 system equipped with standard metal effusion and dopant cells and a showerhead injector for NH$_3$. After appropriate wafer and surface preparation, an AlN nucleation layer was grown [22]. The growth was then paused to change the substrate temperature to 850 °C and to adjust cell temperatures for the AlGaN conditions (Al BEP = $6.44 \times 10^{-8}$ Torr, Ga BEP = $8.59 \times 10^{-8}$ Torr, In BEP = $5 \times 10^{-8}$ Torr).

During this time, a small In flux (BEP = 5×10$^{-8}$ Torr) remained to promote a smooth surface. Then a 1 nm heavily doped Al$_{0.5}$Ga$_{0.5}$N layer ([Si] = 2.5×10$^{20}$ cm$^{-3}$) was grown, followed by a 4 nm UID spacer layer of the same composition. This was followed by another short pause to change the Si cell temperature. Finally, the graded Al$_x$Ga$_{1-x}$N channel was grown with a constant V/III ratio by changing both the Al and Ga cell temperatures simultaneously to ensure constant growth rates and uniform Si doping throughout the channel. The surface displayed a smooth surface morphology characterized using atomic force microscopy (AFM) with a root mean square (rms) roughness of 0.134 nm (Fig. 3(c)). The layer thicknesses and Al composition were confirmed using high resolution X-ray diffraction (HR-XRD) (Bruker XRD) measurements, and fully strained growth was confirmed using XRD-reciprocal space mapping (RSM) measurements shown in Fig. 3(a) and 3(b).

The process flow started with V/Al/Ti/Au (20nm/80nm/20nm/80nm) deposition by e-beam evaporation followed by rapid thermal annealing in N$_2$ at 850 °C for 2 minutes. Cl$_2$-based induced couple plasma-reactive ion etch was used for mesa isolation. Ni/Au (30 nm/100 nm) gate metal layers were patterned by optical and e-beam lithography and deposited using e-beam evaporation.

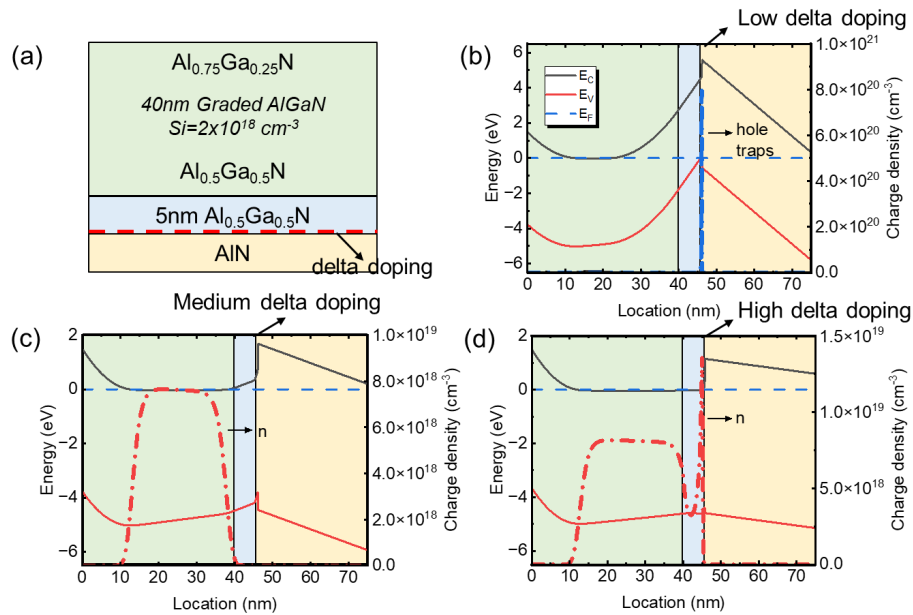

Fig. 2. Schematic and energy band diagrams [21] of the AlGaN PolFET structure. (a) schematic and epitaxial layer structure. Energy band diagrams show three cases with (b) no delta doping (c) moderate delta doping (2.5 x 10$^{20}$ cm$^{-3}$) and (d) high delta doping (4 x 10$^{20}$ cm$^{-3}$) as back barrier.

## III. Results and analysis

Sheet charge density of $1.86 \times 10^{13}$ cm$^{-2}$, electron mobility of 76 cm$^2$/V.s, and sheet resistivity of 4.38 kΩ/□ were extracted from Hall measurements on Van-der-Pauw structures. Capacitance-voltage profiling was used to extract the depth dependence of the apparent charge density (Fig. 4(c)). The apparent carrier density matches well with the simulated one. The bottom of the channel is at the position predicted (~ 40 nm) and no parasitic channel is seen at the AlN/AlGaN interface. This suggests that Si doping compensated the negative polarization charge without inducing an additional parasitic electron channel, similar to Figure 2(c).

DC three-terminal characteristics were measured using an Agilent B1500 parameter analyzer. Transfer length measurements (TLM) show that contact resistance and sheet resistance were estimated to be 4.8 Ω.mm and 4.36 kΩ/□ as shown in Fig. 4(a).

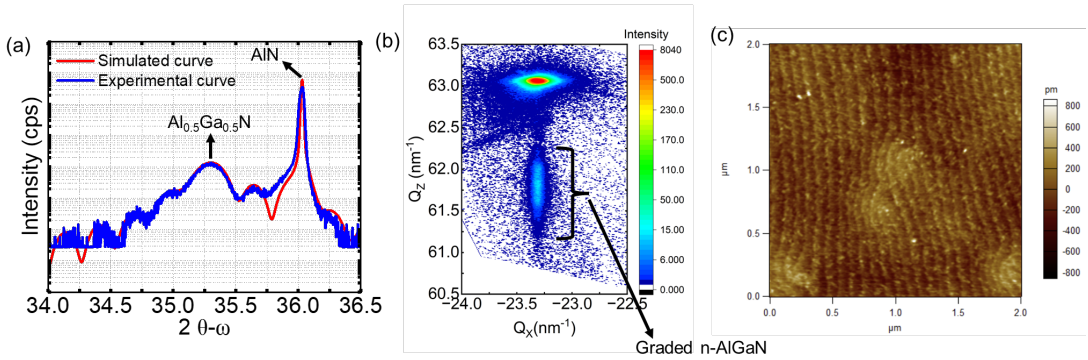

Fig. 3. (a) XRD 2θ-ω scan of PolFET epitaxial layers, (b) XRD reciprocal space mapping data of the epitaxial layer stack, and (c) AFM images on as-grown PolFET surface.

Scaled transistors as shown in Fig. 5(a) with source-drain spacing of 1 μm, gate-drain spacing of 500 nm and gate length of 200 nm are reported here. A maximum current density ($I_{max}$) of over 800 mA/mm was measured at $V_{GS}$ = +2 V and $V_{DS}$ = 20 V as shown in Fig. 5(b). This is among the highest current density values obtained in UWBG AlGaN transistors and is the highest for a graded AlGaN PolFET. A transconductance of approximately 70 mS/mm and threshold voltage of -12 V were extracted from transfer characteristics (Fig. 5(c)). The on-off ratio of the transistors was measured to be ~ $10^2$ limited by the high gate leakage in the samples.

On-wafer small signal measurements were taken using an Agilent network analyzer 8722ES with gate biased at -3.5 V and drain biased at 20 V. From current gain (|h$_{21}$|), maximum unilateral gain (MUG) and maximum stable gain (MSG) versus frequencies as shown in Fig. 6, the device exhibits

current gain cutoff frequency (f$_T$) of 26 GHz and maximum oscillation frequency (f$_{max}$) of 28 GHz. This is an excellent current density and high-frequency performance for high Al-content AlGaN polarization-graded FET and shows the potential of these devices for future RF applications.

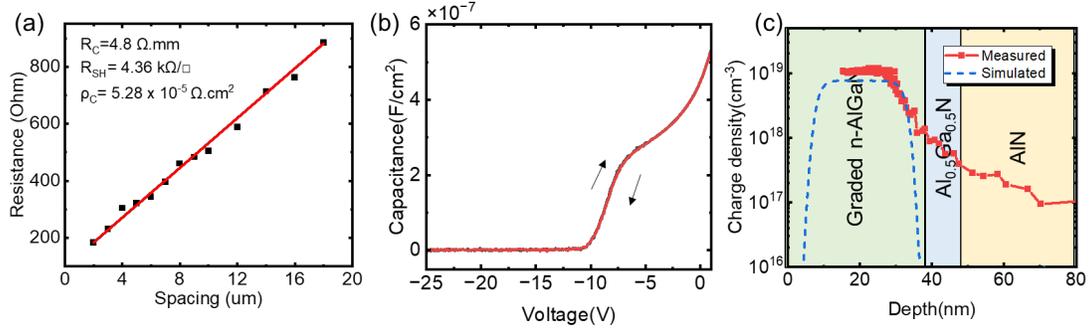

Fig. 4. (a) TLM fitted curve; (b) Double-sweep capacitance-voltage measurements on gate diode of AlGaN PolFET; (c) Extracted charge profile from CV profile and simulated charge profile in Figure 2(c).

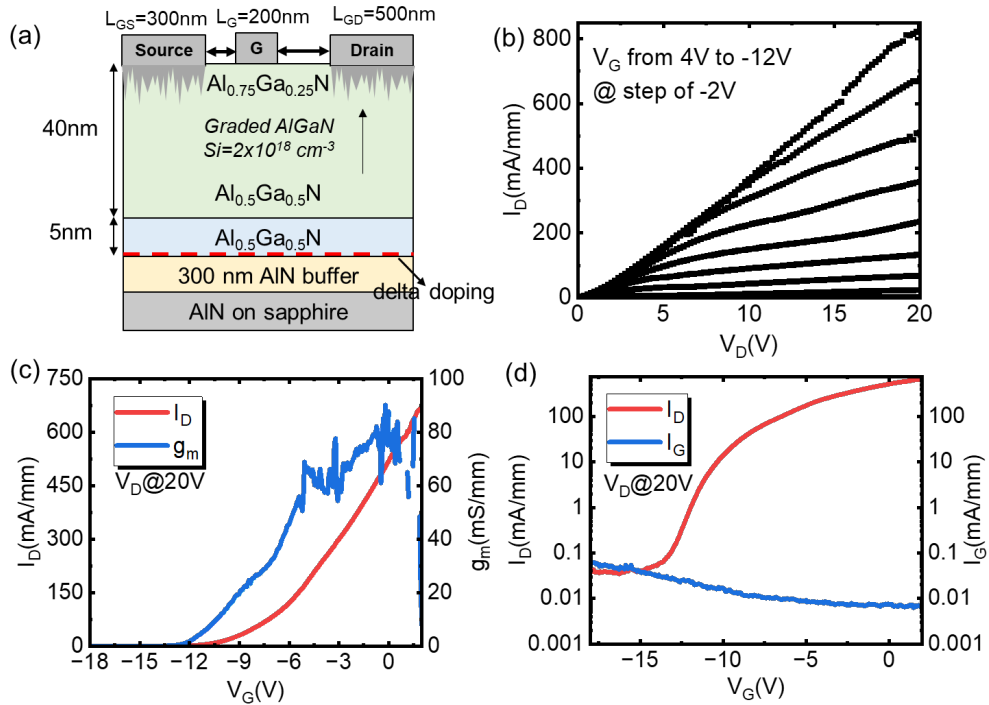

Fig. 5. (a) Schematic of AlGaN PolFET; (b) Output characteristics (c) Transfer characteristics; (d) Drain and gate current (I$_D$ and I$_G$) vs. V$_G$ in log scale.

To further investigate the high-frequency behavior in this device, small signal modeling was done [23,24]. The parasitic and intrinsic circuit elements in the small-signal equivalent circuit model [23] were extracted using S-parameters under different bias conditions, and the parasitic elements and intrinsic elements in the circuit are summarized in Table I. We find that the parasitic delay ($\tau_p$) associated with extrinsic resistances, and transit delay ($\tau_{tran}$) associated with the intrinsic gate-source capacitance [25], can be calculated to be 3.6 ps and 3 ps, respectively. The parasitic delay is mainly attributed to the high contact resistance (4.8 Ω.mm), while the transit delay results from the insufficient scaling of gate length. Future work will focus on improving contacts through integration of a reverse graded contact layer [7] with PolFETs and further scaling of device dimensions.

To thermally characterize the ultra-thin AlGaN PolFETs and compare it to state-of-the-art AlGaN HEMT with much thicker AlGaN buffer layer, gate-resistance thermometry (GRT) was employed. This technique relies on monitoring the temperature-dependent change in the gate metal resistance [26]. Figure 7(c) shows the GRT layout used in this study. A calibration step was first performed to extract the temperature coefficient of resistance (TCR), after which the calibrated relationship between resistance and temperature was used to quantify the device temperature rise under various electrical bias conditions. This approach enables direct evaluation of self-heating during operation.

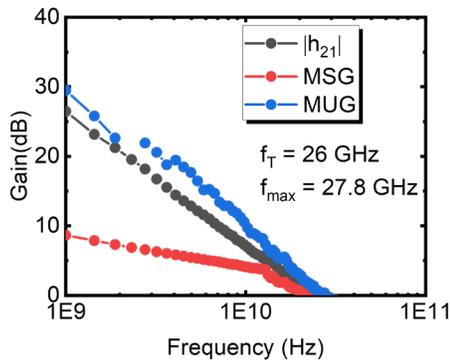

| Intrinsic parameters | |
|---|---|
| $C_{gd}$ | 6.2 x 10⁻¹⁴ F/mm |
| $C_{gs}$ | 1.1 x 10⁻¹³ F/mm |
| $g_{m,int}$ | 90 mS/mm |
| $R_{ds}$ | 90 Ω.mm |
| **Extrinsic parameters** | |
| $R_s$ | 7.404 Ω.mm |
| $R_d$ | 14.82 Ω.mm |
| $R_g$ | 10 Ω/mm |
| **Delays** | |
| $\tau_{tran}$ | 3 ps |
| $\tau_p$ | 3.6 ps |

Fig.6 Extracted current gain, MUG and MSG vs. frequencies.

Table I Elements in small signal modeling

Using the extracted TCR, the temperature rise in each device was measured as a function of applied drain bias. Figure 8 shows the temperature rise versus power density along with linear fits to the data. Thermal resistance values were extracted from these slopes to compare the performance of the two devices. The PolFET device exhibited a thermal resistance of 12 K·mm/W, whereas the HEMT device showed a significantly higher value of 30 K·mm/W. The lower thermal resistance of the PolFET is attributed primarily to its much thinner AlGaN layers, which reduce thermal conduction resistance, as well as differences in device operation that influence heat generation and dissipation. These results highlight the impact of material thickness and device architecture on thermal performance in AlGaN-based technologies.

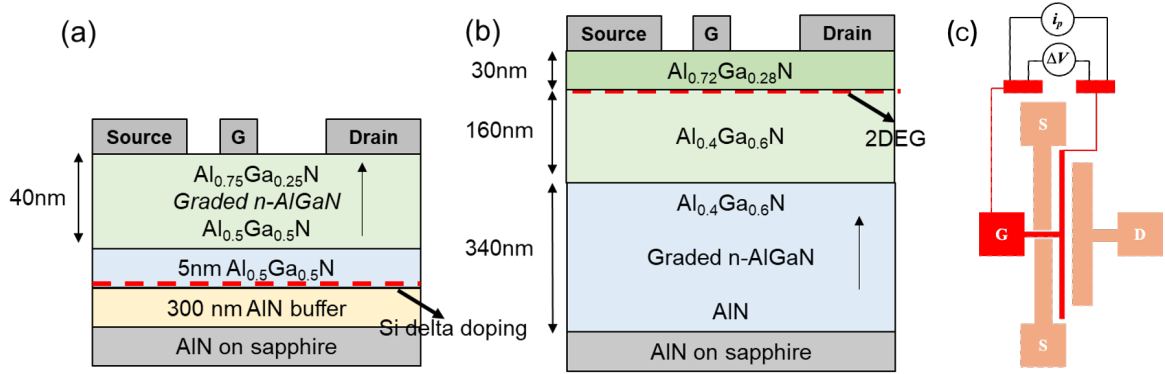

Fig.7 Schematic diagram of (a) AlGaN PolFET with ultra-thin delta-doped back barrier, (b) AlGaN HEMT with thick buffer layer, (c) Device layout and configuration for four-terminal sensing to measure gate resistance.

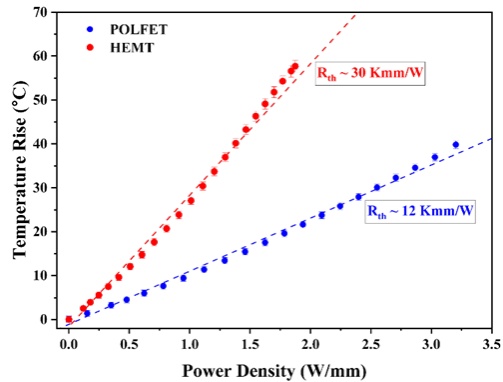

Fig. 8 Temperature rises across the gate metal for different power density for the HEMT and PolFET devices.

### III. Conclusion

This first demonstrated UWBG AlGaN PolFET showed high current density over 800 mA/mm and $f_T/f_{max}$ of 26 GHz/28 GHz, representing state-of-the-art results for UWBG AlGaN transistors. Furthermore, thermal characterization of the ultra-thin PolFET reveals superior thermal performance compared to state-of-the-art thick AlGaN HEMT. These results were achieved using the thinnest epitaxial AlGaN channel layer design reported to date, enabled by a delta-doped interface that can compensate for negative polarization charge. The ultra-thin pseudomorphic AlGaN layers preserve strong carrier confinement and offer improved thermal management in future device generations. These innovations establish a new methodology for AlGaN channel transistors epitaxial layer designs. The benchmarking plots compare well with the state-of-the-art abrupt heterostructures, but the graded structure has important advantages in the integration of reverse-graded ohmic contacts without a barrier (as shown in [27]), improved linearity, and better thermal management. Overall, the ultra-thin epitaxial design concepts demonstrated here represent a significant step toward enabling UWBG AlGaN for future RF and mm-wave electronics.

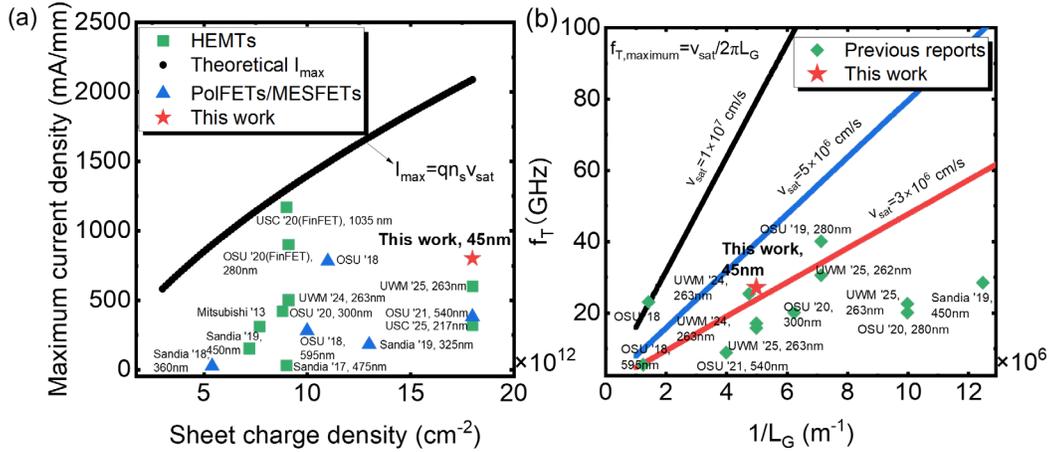

Fig. 9 (a) Benchmarking plot for (a) maximum current density ($I_{max}$) versus sheet charge density ($n_s$) compared to theoretical maximum current density, and (b) current gain cutoff frequency ($f_T$) versus 1/gate length ($L_G$) compared to theoretical maximum current gain cutoff frequency.